# Dual Optical Marker Raman Characterization of Strained GaN-channels on AlN Using AlN/GaN/AlN Quantum Wells and $^{15}$N Isotopes

Meng Qi[1], Guowang Li[1], Vladimir Protasenko[1], Pei Zhao[1], Jai Verma[1], Bo Song[1], Satyaki Ganguly[1], Mingda Zhu[1], Zongyang Hu[1], Xiaodong Yan[1], Alexander Mintairov[1,2], Huili Grace Xing[1] and Debdeep Jena[1,a)]

[1] *Electrical Engineering, University of Notre Dame, Notre Dame, Indiana, 46556, USA*

[2] *Ioffe Physico-Technical Institute, Russian Academy of Sciences, Saint Petersburg, 194021, Russia*

**Abstract**

This work shows that the combination of ultrathin highly strained GaN quantum wells embedded in an AlN matrix, with controlled isotopic concentrations of Nitrogen enables a dual marker method for Raman spectroscopy. By combining these techniques, we demonstrate the effectiveness in studying strain in the vertical direction. This technique will enable the precise probing of properties of buried active layers in heterostructures, and can be extended in the future to vertical devices such as those used for optical emitters, and for power electronics.

[a)] Electronic mail: djena@nd.edu;

III-Nitride semiconductors have found applications in high-speed radio-frequency (RF) devices, ultraviolet (UV) and visible light emitters, and high-voltage power electronics.[1,2,3,4,5,6] For high electron mobility transistors (HEMT), high current density and high switching frequency operation can be achieved at high voltages by utilizing polarization-induced two-dimensional electron gas (2DEG) channels.[7,8] The spontaneous and the piezoelectric polarization charges introduced by the strain at heterostructure interfaces control the charge and device performance.[9,10] Currently popular HEMT structures use AlGaN/GaN, InAlN/GaN heterostructures with thick GaN buffer layers of thickness of a few microns below the 2DEG channel. Optical probing of the HEMT channels by Raman and photoluminescence spectroscopy reveal a wealth of information such as strain, junction temperature, etc that serve as valuable diagnostic tools to study and improve the device performance and reliability. However, because of the presence of a thick GaN layer under the channel, the desired optical signal from the active GaN 2DEG channel mixes with that from the underlying thick GaN buffer layer. Therefore, selective characterization of the thin 2DEG channel region with high precision becomes difficult. It requires fine laser focus, which in turn is limited by the diffraction limit of light. Precise and direct characterization of the GaN 2DEG channel can be achieved by introducing optical markers.[11] In this work, we demonstrate dual optical markers for Raman spectroscopy in III-nitride heterostructures in the HEMT configuration. The method will prove useful for probing optoelectronic devices as well. Thin strained GaN quantum well (QW) channels sandwiched between unstrained AlN barriers serve as the first optical marker, and $^{15}$N isotopes in the crystal serve as the second optical marker for Raman spectroscopy.

As the first optical marker, thin strained GaN quantum wells are sandwiched between unstrained AlN barriers as AlN/GaN/AlN double-heterostructure QWs. The QW structure has shown attractive properties[12] such as the presence of a 2DEG whose charge density can be varied by the QW thickness rather than the barrier thickness. The 2DEG in such QWs boost the largest carrier confinement possible in Al(Ga)N/GaN heterostructures – both on the top and back side. Bulk single-crystal AlN substrates offer the highest thermal conductivity possible in the III-Nitride materials. Taking advantage of these factors, AlN/GaN/AlN double heterostructure HEMTs have been demonstrated[13]. Of special importance to this work is that there is a very small volume of GaN in the QW compared to traditional AlGaN/GaN structures. The 2DEG is also in the GaN QW, and is the channel region of the transistor. This feature makes the GaN QW HEMT structure ideally suited as an optical marker because all other regions – the barrier and the underlying buffer have much larger bandgaps. Furthermore, since Nitrogen is the lightest group-V element, its stable isotope $^{15}$N offers a sufficient mass difference and phonon frequency shift to enable an additional marker, as we demonstrate in this study. Similar studies in other III-V semiconductors are difficult because of the heavy group-V elements that cause minor shifts in phonon frequencies: isotopic control of Nitrogen offers a relatively underexplored, yet powerful Raman metrology tool in III-Nitride heterostructure devices.

In this work, we focus on the study of strain. Raman spectroscopy is used as the characterization method to study phonon-related properties[14,15,16] such as strain[17], isotope effect[18], and thermal properties[19]. Raman measurements have also been extensively used for lattice temperature thermography to study the connection between heat and degradation mechanisms in AlGaN/GaN HEMT[20,21,22]. Strain is also the origin of the piezoelectric polarization charges in GaN. In addition, isotopically mixed Ga$^{14}$N$^{15}$N alloy channels have been theoretically predicted to improve the electron saturation velocity[23] in HEMTs. Raman spectroscopy offers a direct characterization method for the isotope contents and its effects on the phonon spectra.

The AlN/GaN/AlN quantum well structures used in this work were grown on metal-polar Al-polar semi-insulting AlN templates on sapphire or metal-polar bulk single-crystal AlN substrates in a Veeco Gen 930 plasma-assisted molecular beam epitaxy (MBE) system. The growth started with a 200 nm unintentionally doped AlN buffer layer, followed by a thin GaN QW layer whose thickness was varied from 10 nm to 28 nm. A 4.5 nm-thick AlN top barrier and a 2.5 nm-thick GaN cap layer were used to cap the QW. Both AlN and GaN were grown under metal-rich conditions as reported previously[13]. Asymmetric (105) X-ray diffraction (XRD) reciprocal space mapping (RSM) was used to extract the strain and relaxation levels in the GaN QWs. HEMT device structures were fabricated with the heterostructure grown on single-crystal bulk AlN substrate with MBE-regrown n+ GaN source and drain regions to reduce the contact resistance – these structures are used for the Raman study presented here.

Raman spectroscopy was performed in a *Witec* system with a 488 nm wavelength excitation laser under $z(x,x)\bar{z} + z(x,y)\bar{z}$ (normal-incident-normal-collection) geometry. Based on the selection rules[15], the three Raman-active phonon modes are: $E_2^H$ (TO-like), $E_2^L$ (TO-like) and $A_1$ (LO). The atomic vibrations of each phosnon mode are sketched in Fig. 1(a). For a phonon mode propagating along the c-axis, the $A_1$ mode is the LO mode and has the highest energy. The two TO-like modes detectable by Raman are the lower energy modes $E_2^L$ where the heavier (Ga/Al) atoms vibrate, and the higher energy mode $E_2^H$ where the lighter (N) atoms vibrate in the c-plane, as shown in Fig 1(a). All the samples in the work with detail structures are provided in Ref. 24. As calibration measurements, the standard phonon modes and peak of Raman shift values for three GaN layers: single crystal bulk, on SiC, and on sapphire are shown in Fig 1(b). Similarly spectra for two types of AlN: AlN on sapphire, and single-crystal bulk AlN are shown in Fig 1(c).

The GaN-on-sapphire and GaN-on-SiC template substrates show different peak positions compared to bulk GaN. This is due to the residual stress. The shifts compared to the bulk GaN $E_2^H = 567.4\,\text{cm}^{-1}$ are $\Delta E_2^H = +3\,\text{cm}^{-1}$ for GaN-on-sapphire, and $\Delta E_2^H = -0.2\,\text{cm}^{-1}$ for GaN-on-SiC. For the LO mode, the shifts compared to the bulk GaN mode at $A_1 = 734\,\text{cm}^{-1}$ are $\Delta A_1 = +1.4\,\text{cm}^{-1}$ for GaN-on-sapphire and $\Delta A_1 = -0.7\,\text{cm}^{-1}$ for and GaN-on-SiC. The phonon frequency shifts indicate compressive residual stress in GaN-on-sapphire samples, and tensile residual stress in GaN-on-SiC samples. In contrast, $\Delta E_2^L$ modes from both template substrates show $\Delta E_2^L = +0.2\,\text{cm}^{-1}$. A similar effect is observed for AlN substrates in Fig 1(c). Compared with the values in bulk single-crystal AlN substrates, the AlN-on-sapphire substrate show $\Delta E_2^H = +1.6\,\text{cm}^{-1}$, $\Delta E_2^L = +0.4\,\text{cm}^{-1}$ and $\Delta A_1 = +0.9\,\text{cm}^{-1}$. From the measurements above, the $E_2^H$ mode has the largest shift in the peak compared to the $E_2^L$ and $A_1$ modes. This originates from the fact that the $E_2^H$ mode represents in-plane vibration of the lighter nitrogen atoms in the GaN lattice, which is more sensitive to the lateral strain/stress. Based on this observation, the $E_2^H$ mode is mainly used to study the strain properties of the AlN/GaN/AlN QWs in the rest of the work.

Fig. 2(a) shows the Raman spectra of GaN QWs of various thicknesses grown on AlN-on-sapphire templates; the layer structure is indicated in the inset of Fig 2(b). Despite the small thickness and volume of GaN channels (10 nm, 16.5 nm, 28 nm), multiple phonon modes of GaN are observed. Because of the optical marker-nature, the phonon modes are from the ultrathin GaN quantum well or channel layer, with possible small contribution from the much thinner cap layer. The $E_2^H$ mode is observed to shift to higher frequencies with decreasing GaN channel thickness. This indicates a higher compressive strain for thinner GaN QWs. The Raman spectra of the underlying AlN-on-sapphire template as well as that of bulk single-crystal GaN [from Fig 1(b)] are shown as controls in Fig 2(a). The $E_g$ vibrational mode of sapphire masks the $A_1$ mode of GaN. Therefore, we track the higher sensitivity $E_2^H$ mode for the QW series, and plot them in Fig. 2(b). The $E_2^H$ modes for all three QW samples have $\Delta E_2^H > +25 \text{ cm}^{-1}$, indicating much higher strain levels than the modest ~+3.0 and -0.2 cm$^{-1}$ shifts for thick GaN-on-sapphire and GaN-on-SiC templates in Fig 1(a). For the thinnest (10 nm) channel GaN QW, the shift is $\Delta E_2^H = +43 \text{ cm}^{-1}$. The linewidth broadens with increasing QW thickness due to the increasing relaxation.

To show the correlation of the phonon modes measured in Raman spectroscopy with the biaxial strain and relaxation, Fig. 3(a) shows the $E_2^H(GaN)$ phonon peak frequency as a function of the in-plane compressive strain. The strain values were extracted using reciprocal space maps (RSM) from high-resolution X-Ray diffraction scans for the QW samples. Fig. 3(b) and (c) show the RSM data for the 16.5 nm and 10 nm GaN QWs. The biaxial compressive strain causes the shift in the phonon frequency. The magnitude of the shift can be estimated from a linear relation between the biaxial strain $e_{xx}$ via two deformation potential constants $a_\lambda$ and $b_\lambda$ using,

$$\Delta\omega = 2a_\lambda \varepsilon_{xx} + b_\lambda \varepsilon_{zz}, \varepsilon_{zz} = q\varepsilon_{xx}, q = -2C_{13}/C_{33}, \qquad (1)$$

where $\Delta\omega$ is the frequency shift of the phonon energy, and $C_{13}$, $C_{33}$ are the elements of the elastic constant tensor. The estimation of phonon frequency strain relation using the linear model is plotted as the solid black curve in Fig. 3(a) with an error range shown as the dashed lines. The measured values are plotted as red dots. They are close, but outside the linear estimation. The $a_\lambda$, $b_\lambda$, and elastic constant values used ($a_\lambda(E_2^H) = -850 \pm 25 \text{ cm}^{-1}$, $b_\lambda(E_2^H) = -920 \pm 60 \text{ cm}^{-1}$, $C_{13} = -106 \pm 20 \text{ GPa}$, $C_{33} = -398 \pm 20 \text{ GPa}$) are from experimental data of bulk GaN from the literature for a small range of biaxial strain values, as indicated in Fig. 3(a)[15,16 25,26]. The prior model is not able to explain the measured Raman shifts for the highly strained GaN QWs, which calls for additional analysis.

To model the dependence of the Raman frequency on the biaxial strain, a simple mass-spring harmonic oscillator system is a good starting point. The oscillation frequency is $\omega = \sqrt{k/m}$, where $k$ is the force constant and $m$ is the mass. Here $k$ is a simplified measure of the bond strength. Based on the fact that $k$ is inversely propotional to the lattice constant $a$ to the first order $k \propto \frac{1}{a}$, we have $\left.\frac{\partial k}{\partial a}\right|_{a=a_o} \propto -\frac{1}{a_o^2}$ and $\Delta k \propto -\frac{\Delta a}{a_o} \times \frac{1}{a_o} = -\frac{\varepsilon}{a_o}$, where $a_o$ is the unstrained lattice constant. Then the $k(\varepsilon) \sim \varepsilon$ relation can be modeled by $k(\varepsilon) = k_o(1 - C\varepsilon)$. The phonon frequency $\omega(\varepsilon)$ and phonon frequency shift $\Delta\omega(\varepsilon)$ due to strain $\varepsilon$ in this simple model is then

$$\omega(\varepsilon) = \sqrt{k(\varepsilon)/\mu} = \omega_o (1 - C\varepsilon)^{\frac{1}{2}} \approx \omega_o - \frac{1}{2}\omega_o C\varepsilon - \frac{1}{4}\omega_o C^2 \varepsilon^2 - \frac{1}{8}\omega_o C^3 \varepsilon^3,$$

$$\Delta\omega(\varepsilon) \approx -\frac{1}{2}\omega_o C\varepsilon \quad (C > 0),$$

(2)

Where $C$ is a constant and $\mu$ is the reduced mass of the basis atoms. Note that $\varepsilon < 0$ here for the compressive strain ins QWs. The linear model in Eqn. 1 is equivalent to the first-order linear frequency shift $-\frac{1}{2}\omega_o C\varepsilon$. The constant C in Eqn. 2 is equivalent to $C = -\frac{2 \times (2a_\lambda + b_\lambda)}{\omega_o} a_\lambda$, if using Eqn. 1. However, the previous model in literature was based on the experimental data limited to a small range of biaxial strains ($-0.2\% < \varepsilon_{xx} < 0\%$). The experimental data form Ref.16 as well as the corresponding linear relation are plotted in Fig. 3(a). The biaxial strain value in this work covers an order of magnitude larger range ($-2.5\% < \varepsilon_{xx} < -1.5\%$). The Raman frequency shift falls out of the prediction of the linear model based on small biaxial strain values, as shown in red curves in Fig. 3(a). The experimental data also shows non-linear beheavior. One can either adjust the linear constants of Eqn. 1 to explain the high frequency shifts, or consider the higher order non-linear terms of Eqn. 2, which is more accurate. However, owing to the simplicity of the model, we do not attempt to do that here, but stress the need for considering the higher order terms in more elaborate compliance-matrix based calculations.

As a second marker, we show the use of isotopes of nitrogen. Specifically, $^{15}$N is introduced as a second optical marker. The more abundant isotope is $^{14}$N; they differ by one neutron, but are electronically identical. In typical III-Nitride crystals, the $^{14}$N: $^{15}$N ratio is extremely large, because of the relative abundance of $^{14}$N during growth. We intentionally introduce controlled amounts of $^{15}$N during MBE growth into the AlN/GaN/AlN QW heterostructures on bulk single-crystal AlN. The structure studied has a similar AlN/GaN/AlN QW heterostructure as discussed before. It was grown on bulk single-crystal AlN substrates. The isotopic ratio was fixed at $^{14}$N : $^{15}$N = 0.46 : 0.54 throughout the epitaxial process as tracked by the residual gas analyzer in the MBE growth chamber. A 4.5 nm-thick Al$^{14}$N$_{0.46}$$^{15}$N$_{0.54}$ layer served as the electron barrier and a 28 nm-thick Ga$^{14}$N$_{0.46}$$^{15}$N$_{0.54}$ served as the channel, as shown in Fig. 4.

As a control sample, a Ga$^{14}$N QW structure on bulk AlN substrate was grown without the $^{15}$N isotope. In Fig. 4(a), we show the $E_2^H$ peaks for the GaN layer for a control bulk single crystal GaN sample, and the isotope and the non-isotopic QW samples on bulk AlN. Note that a similar phonon peak position was observed for the 28 nm GaN QW samples on the AlN-on-sapphire template (Fig. 2(b)), indicating repeatable growth conditions that helps distinguish the effect of the isotope clearly. The $E_2^H$ and $A_1$ modes were used to study the isotope effect in the GaN channel and the AlN layers. In Fig. 4(a) and (b), for the Ga$^{14}$N$^{15}$N channel, the $E_2^H$ and $A_1$ phonon mode peaks show downward shifts of $\Delta E_2^H = -11.9 \pm 0.28 \, \text{cm}^{-1}$ and $\Delta A_1 = -14 \pm 0.49 \, \text{cm}^{-1}$ compared to the non-isotopic Ga$^{14}$N QW. As discussed earlier, the phonon frequency shift due to isotopic nitrogen can be calculated to first order from the fact that the $E_2^H$ and $A_1$ phonon frequencies are proportional to the square root of the reduced mass of the basis (Ga+N) from the simple harmonic oscillator picture:

$$\frac{\omega(Ga^{14}N_{0.46}{}^{15}N_{0.54})}{\omega(Ga^{14}N)} = \sqrt{\frac{\mu(Ga^{14}N)}{\mu(Ga^{14}N_{0.46}{}^{15}N_{0.54})}} \approx 0.9844, \quad (3)$$

where the reduced mass μ is $\mu^{-1}(GaN) = \mu^{-1}(Ga) + \mu^{-1}(N)$ using the actual percentages of $^{14}N$ and $^{15}N$ in the structures. The estimate yields $\Delta E_2^H(\text{calc}) \sim -9.31 \text{ cm}^{-1}$ and $\Delta A_1(\text{calc}) \sim -11.75 \text{ cm}^{-1}$. The measured values of $\Delta E_2^H = -11.9 \pm 0.28 \text{ cm}^{-1}$ and $\Delta A_1 = -14 \pm 0.49 \text{ cm}^{-1}$ are larger, but close considering the simplicity of the model. In the simplified spring oscillator model, the spatial ordering of $^{14}N$ and $^{15}N$ is assumed to be perfect, with alternating $^{14}N$ and $^{15}N$ in the lattice. The real crystal has an effective isotope alloy disorder because of the statistical randomness in the lattice sites where $^{14}N$ and $^{15}N$ are incorporated. Similar additional frequency shifts in isotopic mixture thick GaN layers has been reported[18]. The additional shift can be related to the spatial disorder using perturbation theory[27,28]. The shift in the peaks makes the isotope containing nitride layer a very sensitive marker. It can be incorporated in layers buried deep within heterostructures and still be probed with fidelity. This is especially well suited for the probing of optical devices such as lasers where the the active regions are far from the surface, and also in vertical devices for power electronics.

Notice in Fig. 4(a) and 4(b) that the FWHM of both modes is larger for the samples that have the isotopic alloy $Ga^{14}N_{0.46}{}^{15}N_{0.54}$ QW compared to the non-isotopic GaN QW, which is turn is larger than the bulk GaN. This indicates a shorter phonon lifetime. The broadening due to isotopic disorder is consistent with expectation because the atomic-scale phonon disorder broadens the optical phonon density of states, which is what is being measured in the Raman spectra. The prediction that this broadening results in a more efficient cooling of hot electrons to increase the drift velocity in GaN transistors remains to

be tested.[23]

As a potential probe of the AlN buffer layer, the effect of the isotopic alloy in the Al$^{14}$N$_{0.46}$$^{15}$N$_{0.54}$ is also picked up in Raman spectroscopy. This is shown in Fig. 4(c) and (d). However, the Raman spectra does pick up the contribution from the isotopic alloy Al$^{14}$N$_{0.46}$$^{15}$N$_{0.54}$ barrier that appears as a weak shoulder to the much stronger peak from the Al$^{14}$N peak. This makes the isotopic effect study of the buffer layers more challenging. The difficulty to separate the peaks from Al$^{14}$N$_{0.46}$$^{15}$N$_{0.54}$ and Al$^{14}$N is because the ~600 μm thick single crystal Al$^{14}$N substrate produces a much stronger signal. One can change the laser focus to enhance the contribution from the isotopic alloy layer, or focus entirely deep into the non-isotopic layer and get estimates of the buffer regions near the channel. Or one can grow a much thicker isotopically mixed buffer to make it directly measurable.

The difference between the Al$^{14}$N$_{0.46}$$^{15}$N$_{0.54}$ shoulder and the Al$^{14}$N peak depends on the relative phonon frequency shift with the atomic mass. The frequency of the Raman mode $\omega_o$ can be estimated by the first-order spring oscillator model as

$$\omega_o = \sqrt{\frac{k}{\mu}} = \sqrt{\frac{k(M_{III} + M_N)}{M_{III} \times M_N}}, \tag{4}$$

Where $k$ is the spring constant of the lattice, $M_{III}$ and $M_N$ are the masses of group III and nitrogen atoms respectively. Then the dependence of the nitrogen atom-induced phonon frequency shift $\Delta\omega/\omega_o$ on the mass of group III atoms in the crystal can be estimated as

$$\frac{\partial \omega}{\partial M_N} = -\frac{k}{2} \times \frac{1}{\sqrt{\frac{k(M_{III} + M_N)}{M_{III} \times M_N}}} \frac{1}{M_N^2}, \frac{\Delta\omega}{\omega_o} = -\frac{1}{2} \times \frac{\Delta M_N / M_N}{1 + \frac{M_N}{M_{III}}}, \tag{5}$$

This means with a heavier group III atom, the phonon frequency shows less relative downward shift $\left|\frac{\Delta\omega}{\omega_o}\right|$ due to the same mass change of isotopic nitrogen atoms. Based on this observation, because the Al atom has smaller mass compared with Ga atom, and the $^{15}$N was

only mixed to 54%, the corresponding $\Delta\omega(AlN)/\omega_o(AlN)$ is small compared with the FWHM of Al$^{14}$N peak.

Furthermore, based on the hypothesis above, $\Delta\omega/\omega_o$ due to isotopic nitrides is expected to follow $\Delta\omega(InN)/\omega_o(InN) > \Delta\omega(GaN)/\omega_o(GaN) > \Delta\omega(AlN)/\omega_o(AlN)$. These experimental studies have not been done and are suggested for the future.

In summary, we have demonstrated the use of strained GaN quantum wells with isotopes as markers for the probing of strain using Raman spectroscopy. We began by calibrating the spectra of typical GaN structures on SiC and Sapphire substrates with bulk single-crystal GaN. We did the same for AlN on sapphire and bulk single-crystal AlN. We used this proper calibration to then demonstrate the measurement of strain in thin GaN QW HEMT channels in AlN. We found that the strain levels are high enough to make them lie outside standard linear approximations of phonon peak shifts. We then demonstrate that isotopes of nitrogen are very effective in offering an additional marker probe of the strain. The phonon peak shifts due to Nitrogen isotopes could be explained from a simple model, and the broadening was indicative of the isotopic alloy disorder. The multiple marker techniques demonstrated here can be applied in several other III-V heterostructures, particularly vertical devices for power electronics, and optoelectronic devices such as LEDs and LASERs, where the active regions are buried far away from the surface.

**Acknowledgement**

This work was supported in part by the Center for Low Energy Systems Technology (LEAST), one of six centers of STARnet, a Semiconductor Research Corporation program sponsored by MARCO and DARPA, and by the Office of Naval Research MURI program monitored by Dr. P. Maki.

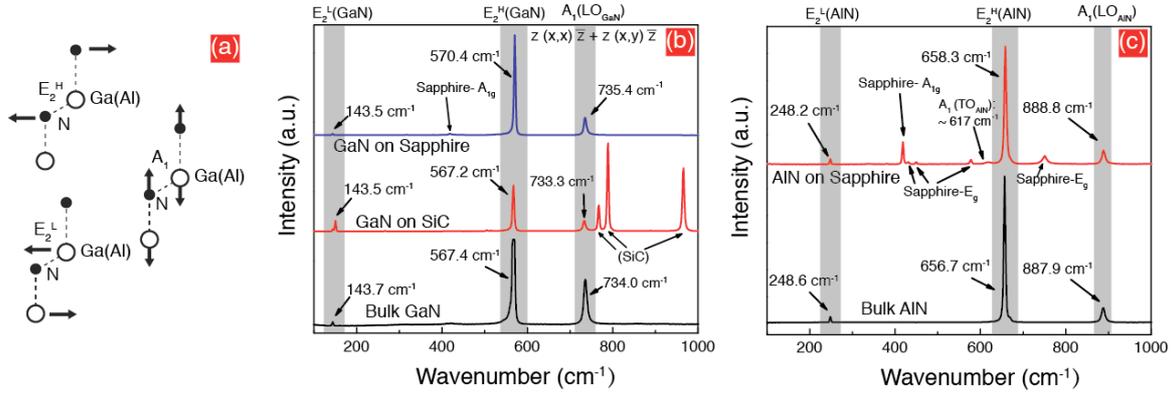

FIG. 1. (a) Raman-active phonon vibration modes $E_2^H$, $E_2^L$ and $A_1$ for GaN and AlN are sketched in terms of the atomic relative vibration, based on the $z(x,x)\bar{z} + z(x,y)\bar{z}$ setup. The measured Raman spectra of (b) three GaN substrates: GaN on sapphire, GaN on SiC, and single-crystal bulk GaN, and (c) two AlN substrates: AlN on sapphire, and bulk AlN. The corresponding modes are labeled. The labeled values for all phonon peak positions were inferred from a Lorentz fit. These control samples have no quantum wells.

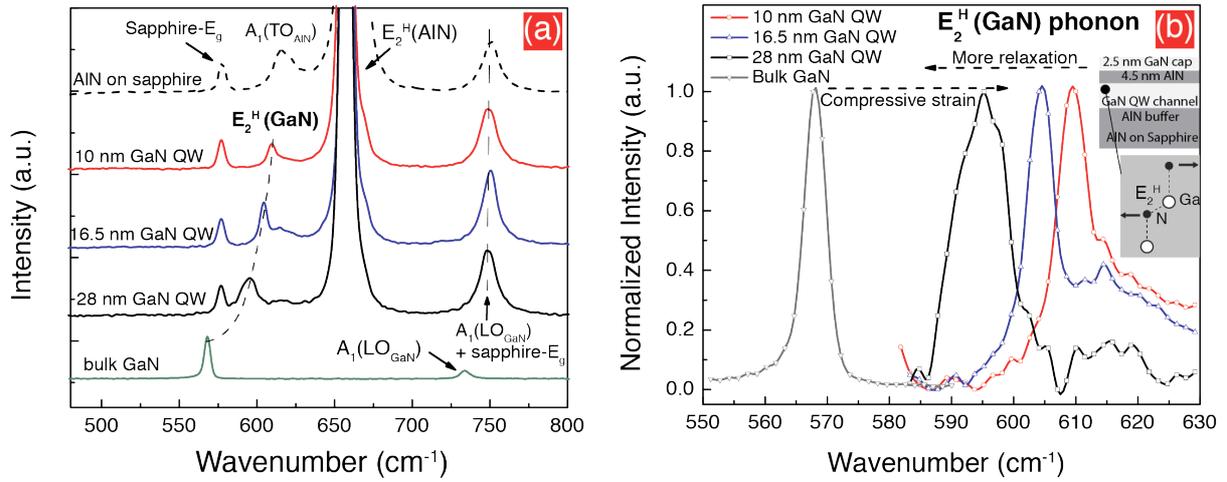

FIG. 2. (a) Measured Raman spectra for AlN/GaN/AlN quantum well structures of thicknesses 10 nm, 16.5 nm and 28 nm on AlN-on-sapphire substrates shown with the spectra from control samples of AlN-on-sapphire (dashed on the top) and bulk GaN (at the bottom). The phonon modes from sapphire are labeled. The spectra of the samples on AlN-on-sapphire substrates are normalized to the sapphire-$E_g$ peak. Note the shift of the $E_2^H$ peak with the thickness of the QW. (b) Normalized $E_2^H$ phonon peak for GaN QW structures of various thicknesses compared with bulk GaN. The strain is highest in the thinnest QW, and increases with the QW thickness. The shifts are all to a $E_2^H$ phonon energy higher than bulk GaN indicating compressive strain. The inset figure in Fig. 2(b) shows the laser focus position in the Raman configuration and the corresponding $E_2^H$ phonon vibration mode measured.

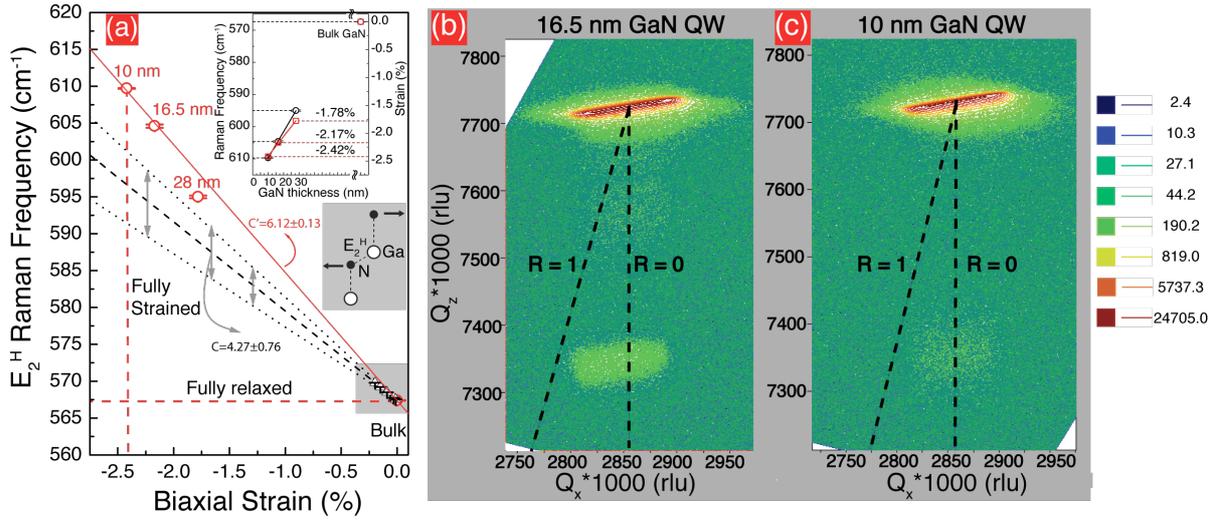

FIG. 3. (a) Raman peak frequencies of the GaN QW $E_2^H$ phonon mode vs the compressive biaxial strain, and vs. GaN QW thickness in the inset. Black lines show the average (solid) and error range (dashed) of calculated values using the linear model of phonon frequency shift vs. biaxial strain. The gray box shows the region of strains of prior experimental data that was used to extract the linear frequency shift coefficients.[16] The prior biaxial strain values are small compared to the biaxial strain in this work. The resulting constant $C$ and $C'$ for the linear dependence are also labeled. The strain values of GaN QWs were extracted by asymmetric (105) reflection reciprocal space X-Ray diffraction maps in (b), (c). The dashed lines represent coherent in-plane strained ($R = 0$) and fully relaxed ($R = 1$) conditions for the GaN QW layer.

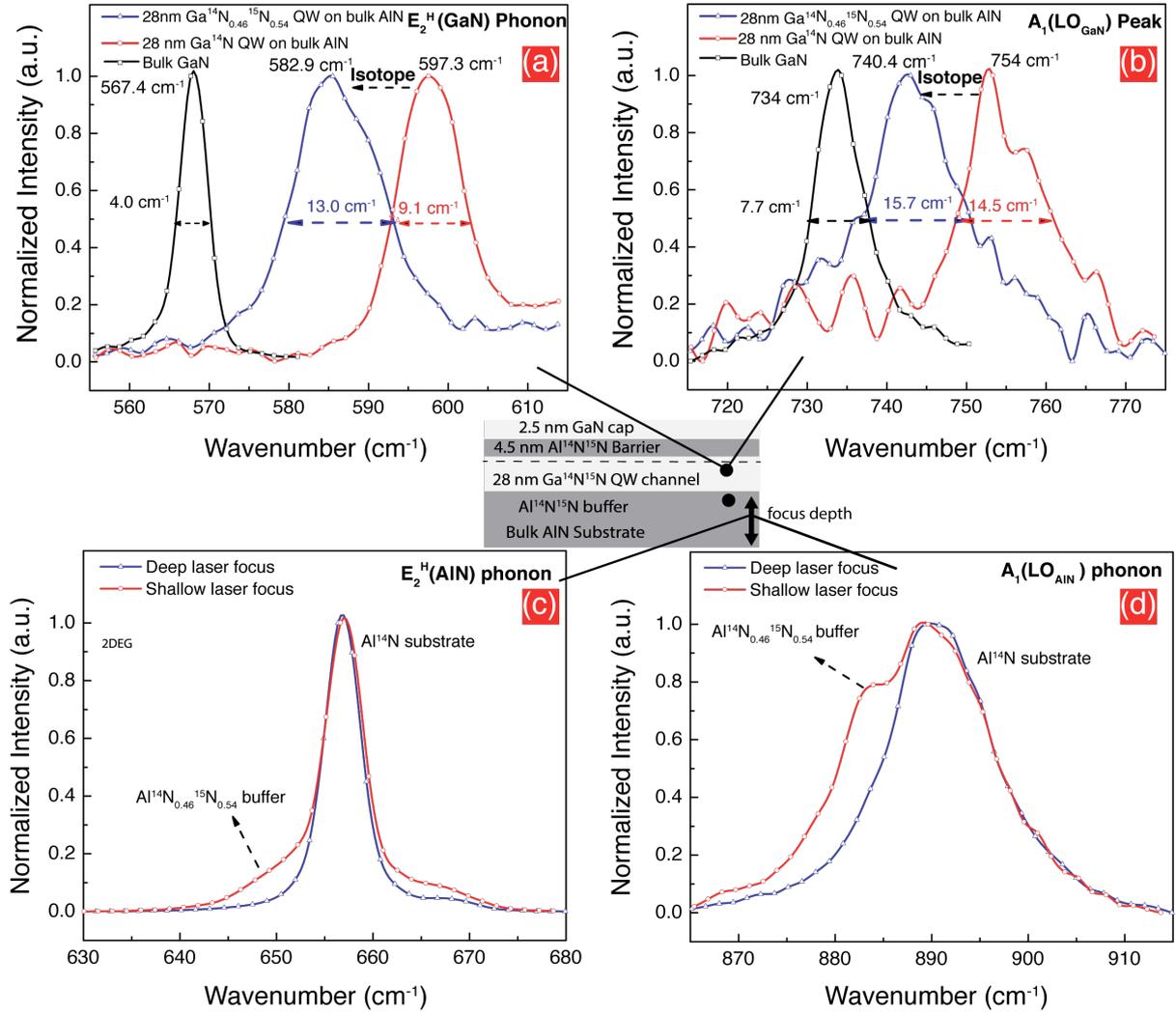

FIG. 4. Measured Raman spectra of 28-nm-thick AlN/GaN/AlN QW structures grown on bulk single-crystal AlN substrates with different GaN channel properties. (a) $E_2^H$ phonon and (b) $A_1$ phonon peak comparison for GaN QW channel layer in 28-nm-thick isotopically pure Ga$^{14}$N QW, isotopically mixed Ga$^{14}$N$_{0.46}$$^{15}$N$_{0.54}$ QW and bulk GaN with normalized intensity. (c) The $E_2^H$ phonon and (b) $A_1(LO)$ phonon comparison for electron barrier of Al$^{14}$N$_{0.46}$$^{15}$N$_{0.54}$ and the bulk Al$^{14}$N substrate in 28-nm-thick Ga$^{14}$N$_{0.46}$$^{15}$N$_{0.54}$ QW FET structure as sketched in the inset. Inset illustrates the two different measurement configurations between (c) and (d) by changing the focal point depth of the excitation laser.